\documentclass[pra,aps,twocolumn,showpacs]{revtex4}
\usepackage{amsmath,graphicx,epsfig}

\def\ket#1{{|#1\rangle}}

\def\ts#1{{_{\mbox{\scriptsize #1}}}}

\def\threej(#1,#2)(#3,#4)(#5,#6){\begin{pmatrix}#1&#3&#5\\#2&#4&#6\end{pmatrix}}
\def\sixj(#1,#2,#3)(#4,#5,#6){\begin{Bmatrix}#1&#2&#3\\#4&#5&#6\end{Bmatrix}}
\def\ninej(#1,#2,#3)(#4,#5,#6)(#7,#8,#9){\begin{Bmatrix}#1&#2&#3\\#4&#5&#6\\#7&#8&#9\end{Bmatrix}}

\newcommand{\prn}[1]{\left(#1\right)}

\newcommand{\sbrk}[1]{\left[#1\right]}

\def\cg(#1,#2)(#3,#4)(#5,#6){{\langle#1,#2,#3,#4}\ket{#5,#6}}

\def\fig_width{3. in} 
\draft

\newlength{\defbaselineskip}
\newcommand{\setlinespacing}[1]%
           {\setlength{\baselineskip}{#1 \defbaselineskip}}

\begin{document}

\title{Nonlinear magneto-optical rotation with frequency-modulated light in the geophysical field range}

\author{V.\ Acosta}
\author{M.\ P.\ Ledbetter}
\author{S.\ M.\ Rochester}
\author{D.\ Budker}\email{budker@socrates.berkeley.edu}
\affiliation{Department of Physics, University of California at
Berkeley, Berkeley, California 94720-7300}

\author{D.\ F.\ Jackson Kimball}
\affiliation{Department of Physics, California State University --
East Bay, Hayward, California 94542-3084}

\author{D. C.\ Hovde}
\affiliation{Southwest Sciences Inc., Cincinnati, Ohio 45244}

\author{W.\ Gawlik}
\author{S.\ Pustelny}
\author{J.\ Zachorowski}
\affiliation{Centrum Bada\'{n} Magnetooptycznych, Instytut Fizyki
im. M. Smoluchowskiego, Uniwersytet Jagiello\'{n}ski, Reymonta 4,
30-059 Krakow, Poland}

\author{V.\ V.\ Yashchuk}
\affiliation{Advanced Light Source Division, Lawrence Berkeley
National Laboratory, Berkeley, California 94720}

\date{\today}


\begin{abstract}
Recent work investigating resonant nonlinear magneto-optical
rotation (NMOR) related to long-lived ($\tau\ts{rel} \sim 1~{\rm
s}$) ground-state atomic coherences has demonstrated potential
magnetometric sensitivities exceeding $10^{-11}~{\rm G/\sqrt{Hz}}$
for small ($\lesssim 1~{\rm \mu G}$) magnetic fields.  In the
present work, NMOR using frequency-modulated light (FM NMOR) is
studied in the regime where the longitudinal magnetic field is in
the geophysical range ($\sim 500~{\rm mG}$), of particular
interest for many applications.  In this regime a splitting of the
FM NMOR resonance due to the nonlinear Zeeman effect is observed.
At sufficiently high light intensities, there is also a splitting
of the FM NMOR resonances due to ac Stark shifts induced by the
optical field, as well as evidence of alignment-to-orientation
conversion type processes. The consequences of these effects for
FM-NMOR-based atomic magnetometry in the geophysical field range
are considered.
\end{abstract}




\maketitle

\section{Introduction}

Sensitive magnetometers are essential tools in many areas of
research, ranging from biology and medicine
\cite{Ham93,Bis03,Bis05} to geophysics \cite{Joh94,Ren88} to tests
of fundamental symmetries
\cite{Mur89,Ber95,Bea00,Rom01,Reg02,Kor02}.  Over the past few
decades, magnetometers based on superconducting quantum
interference devices (SQUIDs) have been the most sensitive
magnetic field sensors \cite{Cla93}. In recent years, however,
significant technical advances have enabled atomic magnetometers
to achieve sensitivities rivaling
\cite{Ale04,Gil01,Wei05,Bud98,Bud00sens,Bud02} and even surpassing
\cite{Kom03} that of most SQUID-based magnetometers. Atomic
magnetometers have the intrinsic advantage of not requiring
cryogenic cooling, and efficient methods for micro-fabrication of
atomic magnetometers (with dimensions $\sim 1~{\rm mm}$) have
recently been developed \cite{Sch04,Balabas05}. Thus atomic
magnetometers offer the possibility of compact, affordable, and
portable ultra-sensitive magnetic sensors.

The atomic magnetometer with the best short-term sensitivity is the
spin-exchange-relaxation-free (SERF) magnetometer whose
magnetometric sensitivity exceeds $10^{-11}~{\rm G/\sqrt{Hz}}$ in
practice and $10^{-13}-10^{-14}~{\rm G/\sqrt{Hz}}$ in principle
\cite{Kom03}.  However, the SERF magnetometer can only achieve
optimum sensitivity at fields ($\lesssim 1~{\rm mG}$) because it
requires the spin-exchange rate to be much greater than the Larmor
precession frequency. Although operation of a SERF magnetometer in
an unshielded environment has been demonstrated \cite{Sel04} by
using a feedback system with three orthogonal Helmholtz coils to
null the ambient magnetic field, it operated far from the limit of
sensitivity due to imperfect cancelation of magnetic noise
(demonstrated sensitivity $\approx 10^{-8}~{\rm G/\sqrt{Hz}}$). In
the present work we demonstrate a technique that enables direct
measurement of geophysical-scale fields with sensitivity $\sim
6\times 10^{-10}~{\rm G/\sqrt{Hz}}$ without requiring feedback
coils. Considerable improvement in sensitivity is expected with
further optimization.

\section{Nonlinear Magneto-optical Rotation with Frequency-Modulated Light (FM NMOR)}

Our approach to atomic magnetometry
\cite{Bud98,Bud00sens,Bud02,Mal04} has focused on the use of
nonlinear magneto-optical rotation (NMOR) \cite{NMOEreview}
related to long-lived ($\tau\ts{rel} \sim 1~{\rm s}$) ground-state
atomic coherences in alkali vapors.  The long coherence times are
obtained by containing room-temperature ($T \sim 20^\circ{\rm C}$)
alkali atoms in evacuated, buffer-gas-free, glass cells with
antirelaxation (paraffin) coating on the inner surface
\cite{Rob58,Bou66,Ale92,Ale02}. The paraffin coating enables
atomic ground-state polarization to survive many thousand wall
collisions \cite{Bou66}, leading to magnetic resonances with
widths of $\lesssim 1~{\rm \mu G}$ \cite{Bud98}.  It has been
demonstrated that this approach offers the possibility of
magnetometric sensitivities on the order of $3\times 10^{-12}~{\rm
G/\sqrt{Hz}}$ in the regime where the Larmor frequency $\Omega_L$
is much less than the relaxation rate of the ground state atomic
polarization $\gamma_\ts{rel}$ \cite{Bud00sens}.

\begin{figure}
\center
\includegraphics[width=3.4 in]{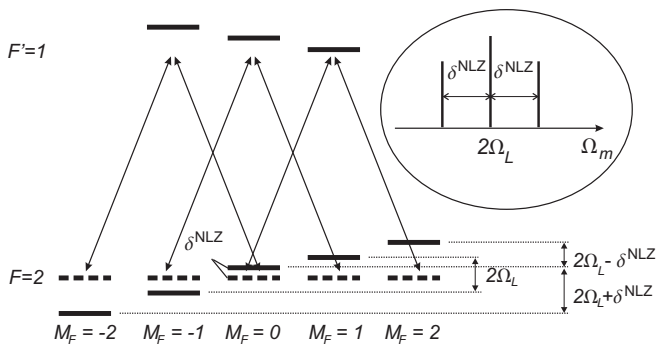}
\caption{Level diagram (not to scale) for the $F=2 \rightarrow
F'=1$ transition investigated in the present work. Dashed bars
represent the unperturbed energies of the ground state Zeeman
sublevels, solid bars represent the energies shifted by a
relatively strong magnetic field.  Linearly polarized light
represented here as a superposition of left- and right-circularly
polarized light (transitions indicated by double-sided arrows)
interacts with three separate coherent superpositions of ground
state Zeeman sublevels, each with $\Delta M_F=2$. Due to the
nonlinear Zeeman effect, the energy difference between the
$M_F=-2,0$ sublevels and the $M_F=0,2$ sublevels differ by $\pm
\delta^{\mathrm{NLZ}}$ from the energy difference between the
$M_F=-1,1$ sublevels [see Eqs.~\eqref{Eq:HFenergy} and
\eqref{Eq:ResFreqs} and surrounding discussion]. This yields three
separate FM NMOR resonances at sufficiently high fields (where
$\delta^{\mathrm{NLZ}} \gtrsim \gamma\ts{rel}$), whose positions
as a function of the modulation frequency $\Omega_m$ are shown in
the inset. Here we ignore coherences and the Zeeman effect in the
excited state, which do not play a prominent role in the effects
discussed in the present work. } \label{Fig:LevelScheme}
\end{figure}

In order to translate the magnetometric sensitivity of NMOR in
paraffin-coated cells to magnetic fields where  $\Omega_L\gg
\gamma\ts{rel}$, the technique of NMOR with frequency-modulated
light (FM NMOR) was developed \cite{Bud02,Mal04}, with estimated
sensitivities better than $10^{-11}\rm{G/\sqrt{Hz}}$. Linearly
polarized, frequency-modulated laser light, propagating in the
direction of the magnetic field, is used for optical pumping and
probing of the ground state atomic polarization. The linearly
polarized light, which can be represented as a coherent
superposition of left- and right-circularly polarized light,
establishes coherences between Zeeman sublevels
(Fig.~\ref{Fig:LevelScheme}). Since in this work the diameter of the
laser beam ($\approx4~{\rm mm}$) is much smaller than the vapor cell
diameter of 3~cm, the atoms that have interacted with the light
subsequently leave the laser beam path and bounce around the cell
thousands of times on average before returning to the light beam in
the appropriate velocity group. The energies of the Zeeman sublevels
are shifted by the magnetic and optical electric field [see
Sec.~\ref{Sec:NLZeemanAndACStark}] causing the relative phase
between the component states of the coherent superposition to evolve
according to the time-dependent Schr\"odinger equation (quantum
beats).  A subsequent light-atom interaction, once the relative
phase between the Zeeman sublevels has evolved, can cause a rotation
of the plane of polarization of the light field (see recent reviews
\cite{NMOEreview,DynamicNMOEreview}).

Since the laser light is frequency modulated, the optical pumping
and probing acquire a periodic time dependence. When the pumping
rate is synchronized with the precession of atomic polarization, a
resonance occurs and the atomic medium is pumped into a polarized
state which rotates around the direction of the magnetic field
(synchronous optical pumping \cite{Bel61a}). Consequently, the
optical properties of the medium are modulated at the quantum-beat
frequency, leading to modulation of the angle of the light
polarization. If the time-dependent optical rotation is measured
at the first harmonic of the modulation frequency $\Omega_m$,
ultra-narrow resonances are observed at near-zero magnetic fields,
and at fields where the modulation frequency $\Omega_m$ is a
subharmonic of one of the quantum-beat frequencies of the system
\cite{Bud02,Mal04,Yas03}. It should be noted that this technique
yields a scalar magnetometer: the precession frequency is
dependent only on the magnitude of the magnetic field and not its
direction.  For light propagation that is not collinear with the
magnetic field, broad, asymmetric resonances are observed at the
same frequency as for propagation parallel to the magnetic field.
At low magnetic fields and low light power, the quantum-beat
frequency for a coherence between two ground-state sublevels
differing in magnetic quantum number by $\Delta M_F$ is $\Delta
M_F \Omega_L = \Delta M_F g_F \mu_B B$, where $g_F$ is the Land\'e
factor for a particular hyperfine component. Here we specialize to
the $^2S_{1/2}$ ground state of the alkali atoms, for which,
neglecting the nuclear magnetic moment, $g_F = \pm g_s/(2I+1)$ for
$F = I\pm 1/2$, where $g_s\approx 2$ is the Land\'e factor for the
electron and $I$ is the nuclear spin. In the range of geophysical
fields and high light powers there are small corrections to the
quantum-beat frequency, as discussed in
Sec.~\ref{Sec:NLZeemanAndACStark}.  In the present work we
concentrate on the signals obtained when $\Omega_m \approx
2\Omega_L$. Such signals are dominated by the precession of the
quadrupole moment (atomic alignment) associated with $\Delta M_F =
2$ coherences, see Refs.~\cite{Yas03,Roc01,NMOEreview}. Higher
order multipole moments were studied in Refs. \cite{Yas03} and
\cite{Pustelny}.

\section{Experimental Apparatus}

\begin{figure}
\center
\includegraphics[width=3.4 in]{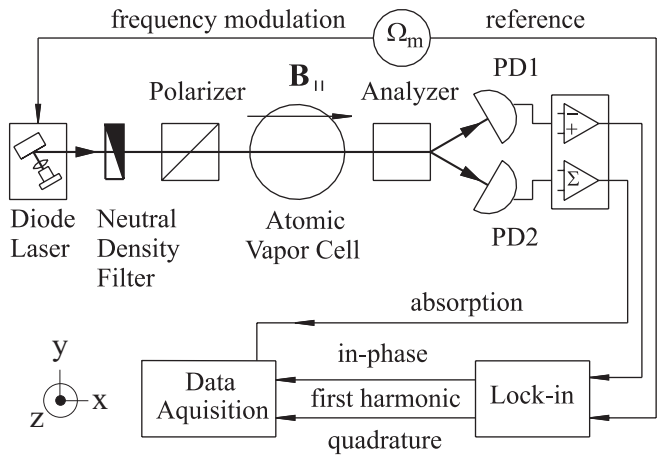}
\caption{Simplified diagram of the FM NMOR setup. The balanced
polarimeter incorporating the polarizer, analyzer and photodiodes
PD1 and PD2 detects signals due to time-dependent optical rotation
of linearly polarized, frequency-modulated light. Not shown are a
set of four-layer magnetic shields and a set of coils to control
the dc magnetic field as well as some
gradients.}\label{Fig:FMNMORsetup}
\end{figure}

Figure \ref{Fig:FMNMORsetup} shows a schematic of the experimental
apparatus. Linearly polarized light propagates along the direction
of the magnetic field, through a spherical, paraffin-coated cell,
with a diameter of 3~cm, containing an isotopically enriched sample
of $^{87}$Rb ($\approx 94\%$ by number, atomic density of $\approx 5
\times 10^9~{\rm atoms/cm^3}$ at $T=20^\circ$C).  The polarization
of the forward-scattered light is monitored using a Rochon
polarizing beam splitter and two photodiodes. The cell is contained
within a four-layer ferromagnetic shield system with shielding
factor $\approx 10^6$ in all directions, described in detail in Ref.
\cite{Yas99}. Inside the innermost shield layer is a set of seven
coils that control the dc magnetic fields ($B_x$, $B_y$, $B_z$) and
first-order gradients ($\partial B_x/\partial x$, $\partial
B_y/\partial y$, $\partial B_z/\partial z$), as well as the
second-order gradient $\partial^2 B_x/\partial x^2$ along the
direction of light propagation ($x$) in which the largest field is
applied. The $B_x$, $B_y$, and $B_z$ coils are rectangular in shape,
matching the shape of the innermost shielding layer so that image
currents in the shields make the coils appear infinitely long,
yielding a field uniform to about a part in $10^4$ over the region
of the cell. The effect of magnetic-field gradients is reduced due
to motional averaging because in the evacuated paraffin-coated cells
the atoms bounce off the walls, traversing the cell many times
between the pump and probe interactions. Nevertheless, we find that
residual gradients contribute to broadening of the FM NMOR resonance
at a level comparable to the effects of power broadening over the
range of magnetic fields studied here.

The frequency of the light from a diode laser (New Focus, Vortex
6000) is modulated with an amplitude of $\approx 120~{\rm MHz}$
(unless stated otherwise) by sinusoidally varying the diode current
at frequencies $\Omega_{m}$ of the order of $2\pi\times 500$~kHz in
this work. The optimum modulation amplitude should be on the order
of the Doppler-broadened width $\approx 300~\mathrm{MHz}$
\cite{Bud02}, but in general is a complicated function of other
parameters such as light power and detuning \cite{Balabas05}. The
central frequency of the laser is tuned $\Delta_0 \approx 200~{\rm
MHz}$ to the low frequency side of the $F=2 \rightarrow F'=1$
hyperfine component of the $^{87}$Rb D1 transition, and stabilized
using a dichroic atomic vapor laser lock \cite{Yas00}. The typical
light power used in these experiments ranged from
$20\thinspace\mu\mathrm{W}$ to 1.0~mW.

\section{Splitting of the FM NMOR Quadrupole ($\Delta M = 2$
coherence) resonance: Nonlinear Zeeman effect and ac Stark shifts}
\label{Sec:NLZeemanAndACStark}

\begin{figure}
  \includegraphics[width=3.4in]{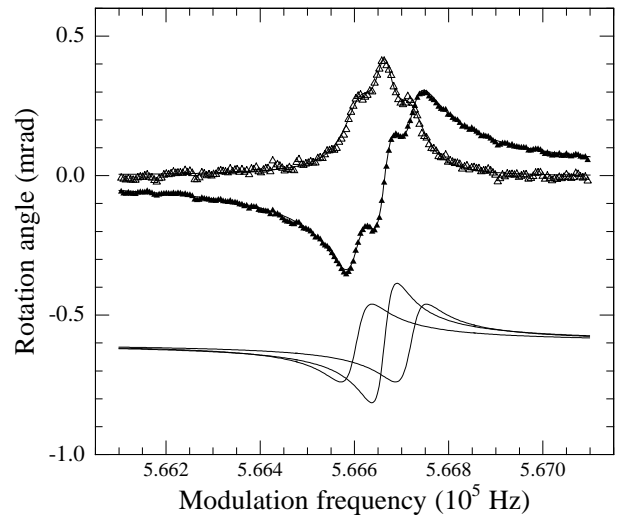}\\
  \caption{Upper traces: Optical rotation signals for B = 405 mG and light power of $95~\mu\mathrm{W}$.
  Solid (hollow) triangles show the in-phase (quadrature) component.  Overlaying the data is a
  fit to three dispersive (absorptive) Lorentzians.  Lower traces: fitted in-phase components of the
  three different $\Delta M_F = 2$ resonances, plotted separately.}\label{Fig:inoutphasesignals}
\end{figure}

The upper traces in Fig.~\ref{Fig:inoutphasesignals} show the
dependence of the amplitude of the time-dependent optical rotation
signal on modulation frequency in the vicinity of the $\Omega_m =
2\Omega_L$ resonance for $B_x=405~{\rm mG}$. Solid symbols
represent the component in-phase with the light modulation, hollow
symbols represent the component out-of-phase (quadrature) with the
light modulation. At the low magnetic fields investigated in
previous FM NMOR studies \cite{Bud02,Mal04,Yas03}, a single,
unsplit, dispersive (absorptive) feature was seen in the in-phase
(quadrature) component. These data demonstrate that in the
geophysical field range, the FM NMOR signals for
$^{87}\mathrm{Rb}$ take on a more complicated shape, the result of
three separate resonances corresponding to the three $\Delta M_F =
2$ Zeeman coherences created in the $F=2$ ground state
(Fig.~\ref{Fig:LevelScheme}).  The quantum beat frequency
associated with each of these coherences has been shifted slightly
due to the combined action of the nonlinear Zeeman effect (mixing
of different ground hyperfine states by the magnetic field) and
the differential ac Stark shift (mixing of the excited state with
the ground state by the optical electric field). Overlaying the
in-phase (quadrature) data in Fig. \ref{Fig:inoutphasesignals} is
a fit to a sum of three dispersive (absorptive) Lorentzians. The
lower traces in Fig. \ref{Fig:inoutphasesignals} show the in-phase
component of each resonance separately.


\begin{figure}
\center
\includegraphics[width=3.4 in]{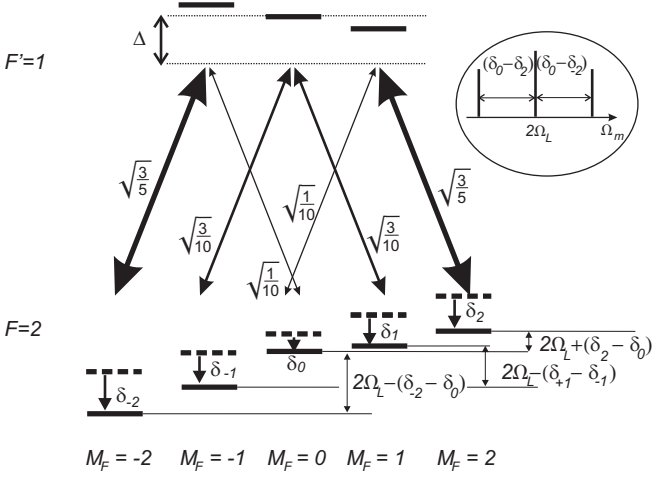}
\caption{Schematic representation (not to scale) of the
differential ac Stark effect for a $F=2 \rightarrow F'=1$
transition. Linearly polarized light, tuned to the low-frequency
side of the atomic resonance ($\Delta < 0$) interacts with three
separate coherent superpositions of ground state Zeeman sublevels
with $\Delta M_F=2$. Dashed bars indicate the energies of the
Zeeman sublevels shifted by the magnetic field
(Fig.~\ref{Fig:LevelScheme}).  The solid bars indicate the
energies of the Zeeman sublevels shifted by the ac Stark effect
due to the optical electric field of sufficiently intense light.
Because the transitions have different dipole moments (indicated
by the thickness of the transition arrows and the numerical values
next to them), the Zeeman sublevels are shifted by different
amounts.  Thus the ac Stark effect can lead to relative shifts by
$\pm (\delta_{\pm 2}-\delta_0)$ of the quantum beat frequencies
for the coherences between the states $\ket{F=2,M_F = 0}$ and
$\ket{F=2,M_F=\pm 2}$ respectively.  The differential shifts due
to the ac Stark effect can either add to or cancel the shifts due
to the nonlinear Zeeman effect, depending on the sign of the
optical detuning $\Delta$.} \label{Fig:LevelSchemeACstark}
\end{figure}

We first consider the low light power limit where the shift in
resonance frequencies of the three coherences is primarily due to
the nonlinear Zeeman effect. Neglecting the contribution of the
nuclear magnetic moment to the Zeeman effect, in sufficiently weak
magnetic fields (where $\Omega_L \ll \Delta\ts{HF}$ and
$\Delta\ts{HF}$ is the energy separation between the hyperfine
levels) the energy $E(F,M_F)$ of a particular ground-state Zeeman
sublevel of an alkali atom with nuclear spin $I=3/2$ is
approximately given by a perturbative expansion of the Breit-Rabi
formula (see, for example, Ref.~\cite{Sobelman}):
\begin{align}
E(F,M_F) \approx E_F + \prn{-1}^F \sbrk{ M_F\Omega_L +
\prn{4-M_F^2}\frac{\Omega_L^2}{\Delta\ts{HF}} }~,
\label{Eq:HFenergy}
\end{align}
where $E_F$ is the energy of the hyperfine level and for $^{87}$Rb
$\Delta_{\rm HF}\approx 6.834~{\rm GHz}$ and we have set $\hbar
=1$. Hence, the quantum-beat frequency $\Omega_{M_F,M_F-2}$ of the
three different $\Delta M_F = 2$ coherences is
\begin{eqnarray}
 \nonumber \Omega_{2,0} &=& 2\Omega_L-\delta^{\rm NLZ}, \\
  \Omega_{1,-1} &=& 2\Omega_L, \label{Eq:ResFreqs}\\
 \nonumber \Omega_{0,-2} &=& 2\Omega_L+\delta^{\rm NLZ},
\end{eqnarray}
where $\delta^{\rm NLZ} = 4\Omega_L^2/\Delta_{\rm HF}$. Note that
the central resonance is unchanged by the nonlinear Zeeman effect
because the states $\ket{F=2,M_F = \pm 1}$ are shifted by the same
amount.  For later convenience, we denote the ``lower'', ``central''
and ``upper'' satellite resonances as those occurring at
$\Omega_{2,0}$, $\Omega_{1,-1}$, and $\Omega_{0,-2}$, respectively.

We now turn to the discussion of the differential ac Stark shift,
shown schematically in Fig. \ref{Fig:LevelSchemeACstark}.  This
effect arises because of 1) differing strengths of the transition
dipole moments from various $F=2$ ground state sublevels to the
$F'=1$ excited state, and 2) differences in optical detuning for the
different hyperfine ground states due to the magnetic field
dependence of the ground state energy levels. A complete treatment
of the ac Stark shifts necessitates a density matrix calculation to
account for optical pumping and the subsequent magnetic field
induced evolution into and out of bright and dark states, and is
complicated by Doppler broadening, the modulation of the light
frequency and the fact that the atoms bounce into and out of the
light. However, some insight can be gained from a heuristic
approach, ignoring these issues and the fact that all laser-induced
transitions are coupled. In this case the ac Stark shift of a two
level system is well approximated by
\begin{equation}\label{Eq:acStarkshift}
    \delta_{M_F} =
    \frac{d_{M_F}^2E^2}{4}\frac{\Delta}{\Delta^2+\Gamma^2/4}
\end{equation}
where $d_{M_F}$ is the relevant dipole matrix element, $E$ is the
optical electric field, $\Delta = \Delta_0+M_F\Omega_L$ is the
detuning from optical resonance and $\Gamma$ is the optical
linewidth.  Neglecting the magnetic-field induced dependence of the
detuning on the magnetic quantum number $M_F$, the shift of the
ground state depends only on the strength of the relevant dipole
matrix element so that $\delta_{M_F} = \delta_{-M_F}$. However, in
the Earth field range, the difference in detuning of the ground
state energy levels is significant compared to the natural
linewidth, yielding an asymmetry in the shifts $\delta_{\pm M_F}$.

Including both the ac Stark shift and the nonlinear Zeeman effect,
the resonance frequencies for the three different $\Delta M_F=2$
coherences are given by
\begin{eqnarray}
 \nonumber \Omega_{2,0} &=& 2\Omega_L-\delta^{\rm NLZ}+(\delta_2-\delta_0), \\
  \Omega_{1,-1} &=& 2\Omega_L+(\delta_{1}-\delta_{-1}), \\
 \nonumber \Omega_{0,-2} &=& 2\Omega_L+\delta^{\rm NLZ}-(\delta_{-2}-\delta_0).
\end{eqnarray}
For detuning to the low frequency side of the optical transition,
the ground state shifts $\delta_{M_F}$ are negative, as indicated in
Fig.~\ref{Fig:LevelSchemeACstark}. Under the assumption that the
detuning does not change too drastically over the $F=2$ manifold,
$\delta_{\pm 2}<\delta_0$ and hence the shift due to the
differential ac Stark effect is in the same direction as the shift
due to the nonlinear Zeeman effect. For detuning to the high
frequency side of the optical resonance, the sign of the
differential ac Stark shift reverses, indicating that it is
primarily off-resonant interaction with the $F'=1$ excited state
responsible for this effect. Since the optical detuning is greater
for the state $\ket{F=2,M_F = -2}$ than $\ket{F=2,M_F = +2}$ by
$4\Omega_L$, $\delta_2<\delta_{-2}$, yielding an asymmetry in the
shift of the upper and lower satellite resonances. Likewise,
$\delta_{1}<\delta_{-1}$ resulting in a small shift of the central
resonance.

\begin{figure}
\includegraphics[width=3.4in]{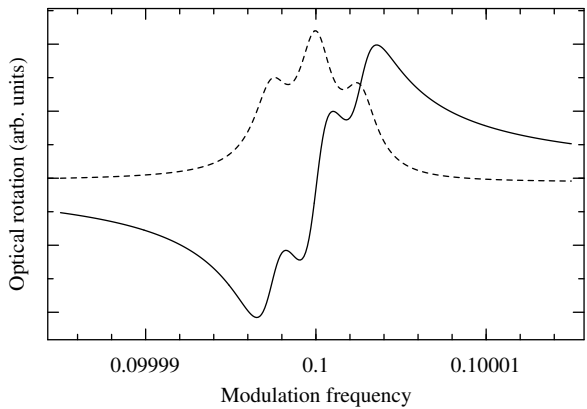}
\caption{Density matrix calculations of in-phase (solid line) and
quadrature (dashed line) optical rotation signals as a function of
modulation frequency for a $F=2\rightarrow F'=1$ transition.
Parameters in units of the natural linewidth $\gamma_0$ are transit
width $\gamma_t=10^{-6}$, modulation depth $=1$, average laser
detuning $=-1$, Rabi frequency (reduced dipole matrix element times
optical electric field) $=3\times 10^{-3}$, Larmor frequency
$\Omega_L=5\times10^{-2}$, and nonlinear Zeeman parameter (shift per
Larmor frequency squared) $=10^{-3}$.}
\label{CalculationBDependence}
\end{figure}

\begin{figure}
\bigskip
\includegraphics[width=3.4 in]{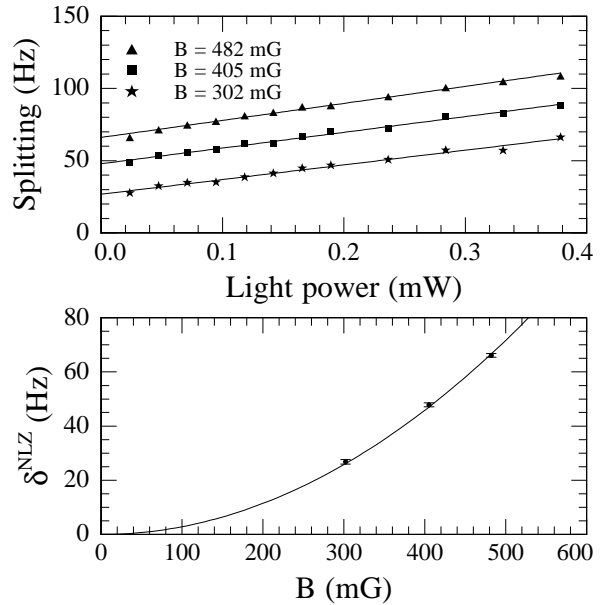}
\caption{Top panel: Total splitting of the FM NMOR resonances
resulting from both the ac Stark shift and the nonlinear Zeeman
effect as a function of light power for different magnetic fields.
Overlaying the data are linear fits.
 Bottom panel: nonlinear Zeeman shift determined by extrapolating to zero
 light power the total splitting $\delta$ in the top panel. Error bars on each
 point are determined by the scatter of the data in the top panel.  The solid line
 is the theoretical prediction, given by $\delta^{\rm NLZ}\approx (\mu_B B)^2/\Delta_{\rm HF}$.}\label{Fig:SplittingVspower}
\end{figure}

A density matrix calculation of FM NMOR on a $F=2\rightarrow F'=1$
transition for atoms in an uncoated cell (where evolution in the
dark can be ignored), neglecting Doppler-broadening, reproduces the
salient features of the experimental data. This numerical
calculation extends our previous analytic calculation \cite{Mal04}
to arbitrary light intensities and angular momenta using the method
of Ref. \cite{Nayak85}. The Hamiltonian for the modulated
light-field interaction and the linear and quadratic Zeeman shifts
is written under the rotating-wave approximation, neglecting terms
counter-rotating at the optical frequency. The density-matrix
evolution equations are then formed, including terms describing
spontaneous decay of the upper state at a rate $\gamma_0$, and atoms
entering and leaving the interaction region (transit relaxation) at
a rate $\gamma_t$. The equations are solved for the Fourier
components of the density matrix using the matrix-continued-fraction
method: a recursion relation for the Fourier components is inverted
as a continued fraction that can be evaluated to any desired
accuracy by truncation (to be described in detail in a future
publication). The resulting optical rotation signal as a function of
modulation frequency is shown in Fig.~\ref{CalculationBDependence}
for parameters similar to those in experimental conditions. The
resonant features have the same shape as seen in
Fig.~\ref{Fig:inoutphasesignals}, including the small asymmetry
between the upper and lower resonances. In addition, the numerically
calculated splittings have a linear dependence on light power, as in
Fig.~\ref{Fig:SplittingVspower}.

It is difficult to distinguish between shifts of the central
resonance and actual drifts of the magnetic field, and hence we
address experimentally only the difference in frequency between the
central and satellite resonances, $\delta_{\rm up} =
\Omega_{0,-2}-\Omega_{1,-1}$ and $ \delta_{\rm lo} =
\Omega_{1,-1}-\Omega_{2,0}$. The top panel of
Fig.~\ref{Fig:SplittingVspower} shows the average of the splitting
between the central resonance and the upper and the lower satellite
resonances $(\delta_{\rm up}+\delta_{\rm lo})/2$ for several
magnetic fields. Overlaying each data set is a linear fit. We find
that the average ac Stark shift is 108~Hz/mW. The offset from zero
of each curve in the top panel of Fig.~\ref{Fig:SplittingVspower} is
due to the nonlinear Zeeman effect. We plot this offset in the
bottom panel of Fig. \ref{Fig:SplittingVspower}. The solid line is
the prediction of Eq.~\eqref{Eq:ResFreqs}, $\delta^{\mathrm{NLZ}}
\approx (\mu_B B)^2/\Delta\ts{HF}$.  A least squares fit to these
data using a purely quadratic model function yields $(287.6\pm
2.6)\times 10^{-6}~\mathrm{Hz/mG}^2$, in agreement with the
theoretical value $286.7\times 10^{-6}~\mathrm{Hz/mG}^2$. We found
that $\delta_{\rm lo}-\delta_{\rm up}$ was approximately linear in
both the magnetic field and light power, and the sign of the
asymmetry was consistent with detuning to the low side of the
optical resonance by more than one linewidth.  At the highest light
power and magnetic field investigated in this work, $\delta_{\rm
lo}-\delta_{\rm up}\approx 20~{\rm Hz}$. We return to the
consequences of this asymmetry for atomic magnetometry in
Section~\ref{Sec:magnetometry}.

\begin{figure}
  \includegraphics[width=3.4in]{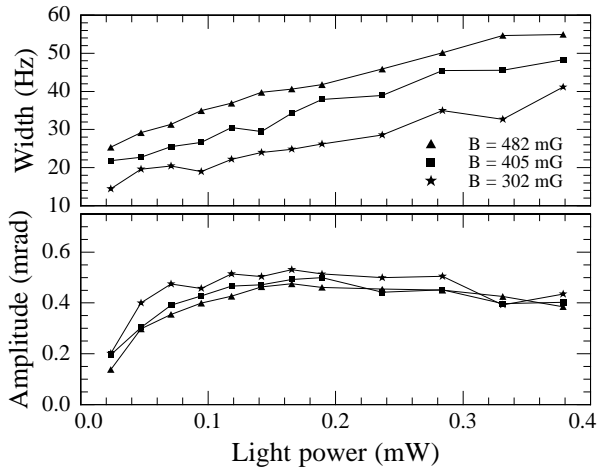}\\
  \caption{Width (top panel, defined in the text) and amplitude (bottom panel) of the central resonance of the in-phase
  component for several different magnetic fields. Points of equal magnetic field are connected to guide the eye.}\label{Fig:ampgamma}
\end{figure}

\section{Light-Power Dependence of FM NMOR Resonance Amplitude}

The top and bottom panels of Fig. \ref{Fig:ampgamma} show the
width and amplitude, respectively, of the in-phase component of
the central FM NMOR resonance as a function of light power for
several magnetic fields. In order to compare with earlier work
\cite{Budker2005}, the width here is defined by the distance from
the center of the dispersive resonance as a function of $\Omega_m$
to the maximum of the rotation angle \cite{widthdefinition}. The
amplitude of the central resonance reaches a maximum of about
0.5~mrad at approximately $150\thinspace\mu\mathrm{W}$,
independent of magnetic field. In addition to power broadening,
there appears to be an extra source of relaxation that increases
with magnetic field. We attribute this effect to high order
magnetic field gradients for which we cannot compensate with the
current experimental setup. At lower magnetic fields $\approx
100~\mathrm{mG}$, by careful zeroing of first order magnetic field
gradients, it was possible to achieve widths extrapolated to zero
power consistent with the best obtained with this cell at 0.15 mG
corresponding to 3.5~Hz \cite{Budker2005}. Finally, we note that
the satellite resonances experience slightly greater light
broadening than the central resonance due to different strengths
of the transition dipole moments.  First order perturbation theory
predicts a ratio of 3:4 between the width of the central resonance
and width of the satellite resonances, in agreement with our
measurements.

\section{Alignment-to-orientation conversion (AOC)}\label{Sec:AOC}
\begin{figure}
  \includegraphics[width=3.4in]{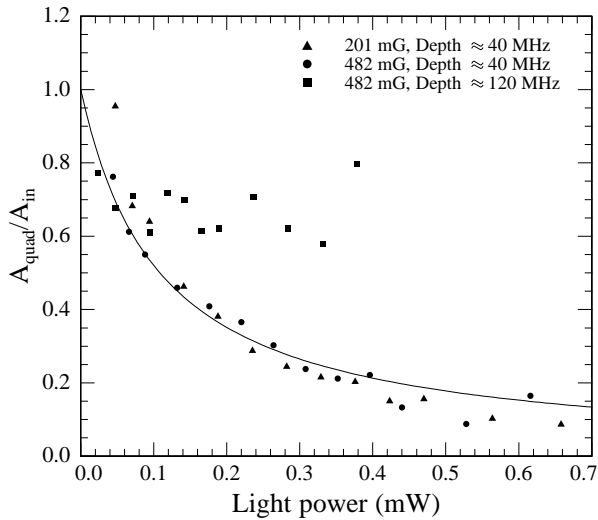}\\
  \caption{Ratio of the in-phase to quadrature amplitudes for the central resonance at several different
  magnetic fields and modulation depths. Overlaying the data for a modulation depth of $\approx 40$~MHz is a
  smooth curve to guide the eye. }\label{Fig:inoutampcomp}
\end{figure}

In the presence of both the magnetic field and the ac Stark shift
due to the optical electric field, the atomic alignment induced by
optical pumping can be converted to an orientation (dipole moment)
parallel to the magnetic field, a process known as
alignment-to-orientation conversion (AOC) \cite{AOCRefs,
Balabas05}. A static orientation along the magnetic field leads to
optical rotation via circular birefringence which is maximal when
the light is tuned to resonance. Thus as the light frequency is
periodically tuned closer and further from the resonance, the
amplitude of rotation due to the static orientation increases and
decreases in phase with the light frequency modulation. To
illustrate the effect of AOC, we plot the ratio of the amplitude
of the in-phase and quadrature components as a function of light
power in Fig. \ref{Fig:inoutampcomp} for modulation depths of
40~MHz and 120~MHz. For a modulation depth of 40~MHz (circles and
triangles), we find that the amplitude of the quadrature component
decreases relative to the in-phase component as the light power
increases, independent of the magnetic field, indicating the
presence of alignment-to-orientation conversion. This effect is
reduced dramatically for data taken with a modulation depth of
120~MHz, consistent with results presented in
Ref.~\cite{Balabas05}.  The simplest explanation for this behavior
is that at higher modulation depths, the light field is, on
average, further from resonance, leading to smaller ac Stark
shifts.  However, the observed ac Stark shifts were similar for
data sets with both large and small modulation depths.  Further
work will be necessary to fully understand AOC with frequency
modulated light.

\section{Atomic magnetometry in the geophysical field
range}\label{Sec:magnetometry}

The sensitivity to magnetic fields is given by
\begin{equation}\label{Eq:sens1}
    \delta B = \left (\frac{d\phi}{dB} \right)^{-1}\delta \phi
\end{equation}
where $\phi$ is the amplitude of the synchronously detected
optical rotation angle and $\delta\phi$ is the angular sensitivity
of the polarimeter. In the present work, we scan the modulation
frequency rather than the magnetic field, so to interpret the
results given here in the context of atomic magnetometry, we write
\begin{equation}\label{Eq:sens2}
    \delta B = \left (\frac{d\phi}{d\Omega_m}\frac{d\Omega_m}{dB} \right)^{-1}\delta
    \phi.
\end{equation}
Here $\Omega_m = 2g_s\mu_B B/(2I+1)\approx \mu_B B$ for
$^{87}\mathrm{Rb}$, $d\phi/d\Omega_m\mid_{2\Omega_L}=A/\gamma$ where
$A$ is the amplitude and $\gamma$ is the width of the resonance, so
that
\begin{equation}\label{Eq:sens3}
    \delta B = \frac{\gamma}{A\mu_B}\delta\phi.
\end{equation}
The shot noise limited angular sensitivity of a polarimeter for
$150\thinspace\mu\mathrm{W}$ of radiation at 795 nm is about
$4\times 10^{-8}\thinspace\mathrm{rad}/\sqrt{\mathrm{Hz}}$. At this
light power $A \approx 5\times 10^{-4}\thinspace\mathrm{rad}$ and
neglecting broadening due to magnetic field gradients, we assume a
width of about $\gamma\approx 2 \pi\times 10\thinspace\mathrm{Hz}$.
Inserting these numbers into Eq. (\ref{Eq:sens3}) we get
\begin{equation}\label{Eq:sens4}
    \delta B \approx 6\times
    10^{-10}\thinspace\mathrm{G}/\sqrt{\mathrm{Hz}}.
\end{equation}
This falls somewhat short of the sensitivity estimate of
$10^{-11}~{\rm G}/\sqrt{\rm Hz}$ \cite{Bud02} derived from
measurements performed at roughly 1~mG. The largest factor
contributing to the reduced sensitivity is the use of a 3~cm
diameter cell rather than a 10~cm cell which yielded a larger
optical depth and FM NMOR resonance widths of about 1~Hz. Additional
contributions to the higher sensitivity reported in Ref.
\cite{Bud02} come from the fact that the splitting of the FM NMOR
resonance was much less than its width. Hence three different
$\Delta M_F = 2$ coherences contributed to the slope of the optical
rotation signal as a function of magnetic field. While there is
little that can be done about the separation of the resonance
frequencies for a given atom at high magnetic field, we anticipate
considerable improvement with further optimization of parameters
such as light power, modulation depth, detuning of the central
optical frequency, cell size and temperature. Unfortunately,
environmental magnetic noise (the current sources used to generate
the magnetic field are only stable to about a part in $10^6$)
prevents a demonstration of sensitivity in the Earth-field range
beyond the level of about $1\mu {\rm G}/\sqrt{\rm Hz}$.

We now briefly address the issue of the shift of the central
resonance due to the magnetic-field induced asymmetry in optical
detuning.  We first make the assumption that we are detuned to the
low frequency side of the optical resonance by more than one
linewidth, and that $4\Omega_L$ is much less than the optical
linewidth.  Under these conditions $\delta_2-\delta_{-2}\approx
4(\delta_1-\delta_{-1})$ because the relevant transition dipole
matrix elements differ by a factor of $\sqrt{2}$ (see
Fig.~\ref{Fig:LevelSchemeACstark}) and the difference in detuning
between the $M_F = \pm 2$ sublevels is twice as great as for the
$M_F = \pm 1$ sublevels. Making use of this fact we see that
$\delta_{\rm up}-\delta_{\rm lo} \approx 2(\delta_1-\delta_{-1}).$
Hence the central resonance is shifted from $2\Omega_L$ by only 10
Hz out of 700~kHz at the highest magnetic field and light power in
this work. In principle, it should be possible to account for this
in a practical device.

\section{Conclusions}

We have performed an experimental investigation of nonlinear
magneto-optical rotation with frequency-modulated light (FM NMOR)
resonances in the geophysical field range. We have observed a
splitting of the usual quadrupole FM NMOR resonance into three
distinct resonances by the nonlinear Zeeman effect as well as by
the ac Stark shift. Our measurements of the nonlinear Zeeman
effect are in good agreement with theoretical predictions.  The
combined effects of the ac Stark shift and the linear Zeeman
effect lead to an asymmetry in the splitting of the FM NMOR
resonance. Evidence for alignment-to-orientation conversion has
been presented. Finally, based on the measurements described here,
we estimate a sensitivity to magnetic fields based on FM NMOR
resonances of $\sim 6 \times 10^{-10}~{\rm G}/\sqrt{\rm Hz}$ and
we anticipate significant improvement with further optimization.

\acknowledgments The authors would like to thank A.I. Okunovich and
J. Higbie for useful discussions.  This work is supported by DOD
MURI grant \# N-00014-05-1-0406, by NSF grant \# INT-0338426, by DoD
and NASA STTR programs by the European Social Fund and by the Polish
Ministry of Education and Science.


\begin{thebibliography}{99}
\bibitem{widthdefinition} In work reported in Ref. \cite{Budker2005}, the magnetic field,
rather than the modulation frequency, was scanned and the width
was defined by $g_s \mu_B/(2I+1) \Delta B$ where $\Delta B$ was
the distance between the extrema of the dispersive resonance.  In
this work we scan the modulation frequency and the resonance
occurs at, $\Omega_m = 2 g_s \mu_B/(2I+1) B$, and hence we quote
the distance from the center to the maximum of the dispersive
signal as a function of $\Omega_m$.
\end{thebibliography}
\end{document}